# Microgrid Resilience: A Holistic and Context-Aware Resilience Metric

Sakshi Mishra[a], Ted Kwasnik[a], Kate Anderson[a*],
[a]National Renewable Energy Laboratory, Golden, USA

Emails: Sakshi.M@outlook.com, Ted.Kwasnik@nrel.gov, Kate.Anderson@nrel.gov
*corresponding author

***Abstract***— Microgrids present an effective solution for boosting the resilience of distribution systems because they serve as a backup power source when the utility grid's operations are interrupted due to either high-probability, low-impact events like a component failure or low-probability, high-impact events such as a natural disaster or cyberattack. However, the degree to which a microgrid can defend, adapt, and restore normal operation depends on various factors including the type and severity of events to which a microgrid is subjected. These factors, in turn, are dependent on the geographical location of the deployed microgrid as well as the risk profile of the site where the microgrid is operating. In this work, we attempt to capture this multi-dimensional interplay of factors in quantifying the ability of the microgrid to be resilient in these varying aspects. This paper proposes a customized site-specific quantification of the resilience strength for the individual microgrid's capability to absorb, restore, and adapt while maintaining power to critical loads when a low-probability, high-impact event occurs. This ability is measured as a *context-aware resilience metric*. A case study illustrates the key elements of the proposed integrated analytical approach.

***Keywords* — *Microgrid, Resilience, Resilience-metric, Resilience quantification, Risk assessment, System operation***

## I. Introduction

### *A. Background and Literature Review*

Energy systems are undergoing a major transformation as countries around the world work towards meeting their decarbonization and resilience targets. Shaping this transformation are factors like renewable energy technologies (distributed energy resources and energy storage) as well as the need for a continuous and dependable electricity supply. Energy systems have long been categorized as a part of a country's critical infrastructure because of the reliance of the modern economy on electricity [1]. In this environment, the ability to maintain power to critical loads in the event of a natural disaster or a cyber-incident has become a major concern. Microgrids represent a subcategory of power grids. When connected to the utility grid, they are designed to sustain critical loads in case of a grid outage [2]. When designed to operate in isolation in locations like remote islands, microgrids are the only source of electricity, and the resilience of the microgrid becomes all the more important.

This work seeks to develop a method to quantify the degree or the strength of the *resilience* of the microgrid that will inform design decisions. Metrics to quantify reliability differ from the metrics needed to quantify resilience. Reliability can be defined as the ability of the system to deliver the quantity and quality of electricity required. Reliability is generally measured by standardized metrics such as Customer Average Interruption Duration Index (CAIDI), System Average Interruption Frequency Index (SAIFI), and System Average Interruption Duration Index (SAIDI). These metrics provide a measure of the system's capability to maintain power to loads in a consistent manner. They represent a *binary view* of the system performance, either the system is functional or it has failed. On the other hand, resilieince is concerned with the ability of a system to recover and restore normal operations when a negative event occurs. The National Infrastructure Advisory Council [3] defines critical infrastructure resilience as: "…the ability to reduce the magnitude and/or duration of disruptive events. The effectiveness of a resilient infrastructure or enterprise depends upon its ability to anticipate, absorb, adapt to, and/or rapidly recover from a potentially disruptive event." Our previous work has dealt with a detailed definition and interpretation of the concept of resilience (subsection 2.2 in [4]). Given the differences between reliability and resilience, different metrics are needed to measure and quantify these two related, but distinct concepts.

The concept of resilience has been extensively explored in the context of power systems (i.e. the centralized utility grid). For example, Wang et al [5] and Jufri et al [6] provide a comprehensive review of the research on resilience of power systems during natural disasters. The key strategies for realizing resilience in power systems are reviewed in [7]. Vugrin et al [8] evaluate the effect of resource constraints on the resilience of bulk power systems. An assessment of power systems resilience during hurricanes is conducted in [9]. Fang et al focus on investment optimization in power systems with the objective of building resilience against attacks [10]. Resilience indicators are proposed for the evaluation of energy infrastructure in [11]. A load restoration framework based on distribution automation technology in the context of power systems resilience is presented in [12].

The role and significance of microgrids in enhancing the resilience of power systems has also been studied in the literature to a large degree. A system-level assessment of reliability and resilience provision from microgrids is presented in [13]. A method for load restoration through a microgrid formation strategy is proposed in [14]. Similarly, evaluation of microgrid's ability to serve as a resilience resource for the utility grid is studied in [15] [16] [17]. But the question of a microgrid's *own* resilience in the face of low-probability high-impact events such as natural disasters and cyber-attacks remains relatively unexplored. One example in the literature addresses this question from the specific angle of windstorms [18], but does not comprehensively address other potential hazards. An all-inclusive view should account for various types of events and the multi-dimensional interplay of a microgrid's infrastructural and operational aspects in both physical and cyber domains. This comprehensive study of the problem can be approached either in qualitative or quantitative ways.

Our previous work has presented a qualitative approach to assessing the resilience of microgrids [4]. In the current work presented here, we employ a quantitative approach to assessing microgrid resilience. We present a resilience metric formulation that accounts for various threats and vulnerabilities associated with those threats within a microgrid. We take a probabilistic approach to arrive at a metric that captures the interplay of the *threat*, *vulnerability*, and *vulnerability impacts* within a microgrid. The need and relevance of this work are further established in the following subsection.

### *B. Relevance and Contribution*

Resilient microgrids, when designed with needed capabilities, can effectively provide a reliable and robust supply of backup power, withstand threats, adapt to continually changing circumstances during an event, and trigger either automated or semi-automated restorative actions after the disaster has passed. However, the challenge lies in determining an effective design for a specific case. For example, if a microgrid is situated in a coastal area, close to sea level, then the effective design choice against the *specific threat* of hurricane-induced flood would be to elevate the on-site generation. However, this seemingly straightforward solution doesn't provide the microgrid with resilience against a cyber attack or fire. The design needs to consider and quantify all possible threats, vulnerabilities, and vulnerability impacts in parallel.

It is important to note that designing resilient microgrids requires an interdisciplinary approach considering techno-economic, socio-economic, and socio-technical factors (Figure 1). Furthermore, there are wide variations in the human judgment of the risk and existing vulnerabilities in the system. This work, therefore, couples mathematical risk assessment methods (Monte Carlo simulation) with human expert knowledge about the risk-vulnerability profile of a given site to develop a holistic picture of this problem's solution space in the socio-technical domain.

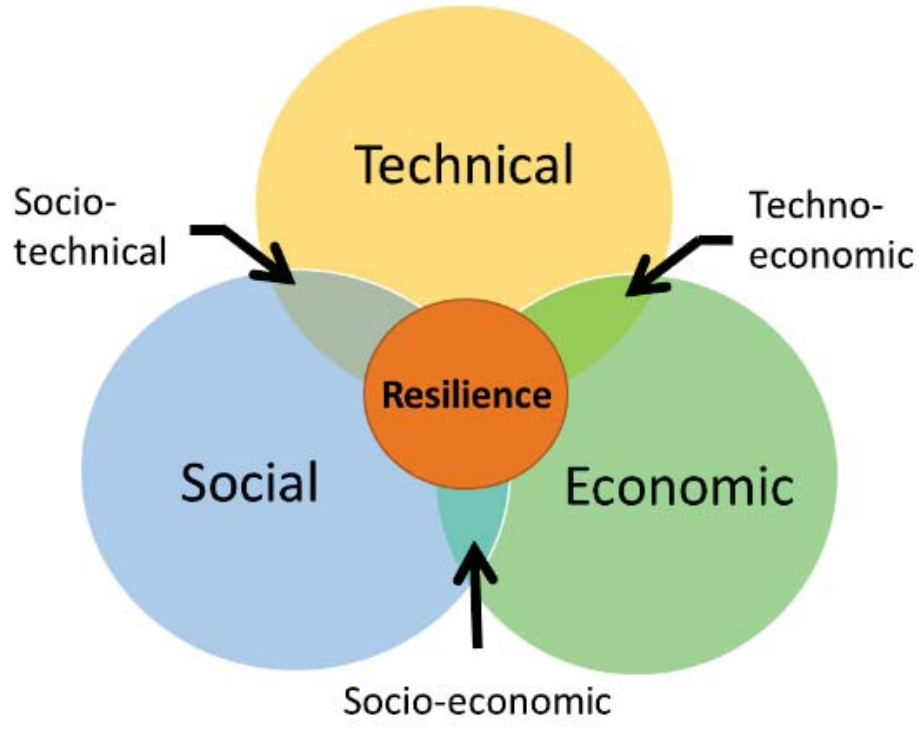

**Figure 1 Resilience - an interdisciplinary problem-space**

The proposed approach is holistic because it considers both the physical and cyber layers of the microgrid, and context-aware because it considers both operational and infrastructural aspects. The specific contributions of the work are as follows:

- We propose a resilience hierarchy for assessing the multi-faceted nature of the microgrid resilience concept.
- We layout the framework for a context-aware and holistic quantitative resilience metric that can be used for assessing the resilience potential of a given microgrid design.
- We demonstrate the proposed framework to determine the resilience baseline of a microgrid through a detailed case study.
- We present an approach to employ the resilience baseline to identify the most effective interventions, which can then be prioritized over other possible interventions, to enhance the resilience capabilities of the microgrid.

### C. Article Structure

Section I laid out the background and motivation for this work and listed our contributions to the research question of microgrid resilience quantification. The rest of the article is organized as follows: Section II defines microgrid resilience and presents our proposed hierarchical relationship between the infrastructural and operational resilience dimensions of the microgrid. A novel framework for microgrid resilience metric calculation is introduced in Section III. Section IV presents a case study based on the proposed framework and Section V concludes with a discussion of future research directions.

## II. WHAT DOES IT MEAN FOR THE *MICROGRID* TO BE RESILIENT?

The United States Presidential Policy Directive (PPD21) [1] defines the term "resilience" to mean "the ability to prepare for and adapt to changing conditions and withstand and recover rapidly from disruptions. Resilience includes the ability to withstand and recover from deliberate attacks, accidents, or naturally occurring threats or incidents." There are many variations of the definitions of *power systems resilience* presented in the literature [19], but most are based on the four phases of anticipating, resisting or absorbing, responding or adapting, and recovering from a disturbance, as shown in Figure 2. For this work, we consider the microgrid's role in each of these phases, and define a 'resilient microgrid' as the ability of the microgrid to:

- Be resistant against the potentially damaging event;
- Be absorptive of the impact of the disturbance introduced by the event, without losing the critical load[1] by having needed redundancies (uninterruptible power supply, for example) and by being operationally adaptive (for example, continuing to serve the critical load by rerouting energy from available generation resources);
- Rapidly restore normal operations after the event has passed through automated or semi-automated operations restoration, or in case of infrastructure damage, through trained personnel capable of operating partially-activated assets in manual operational mode and through resourcefulness in availability of spare parts to repair damaged systems;
- Be infrastructurally adaptive by designing the system with upgraded physical and cyber resilience features to operate in semi-automated mode.

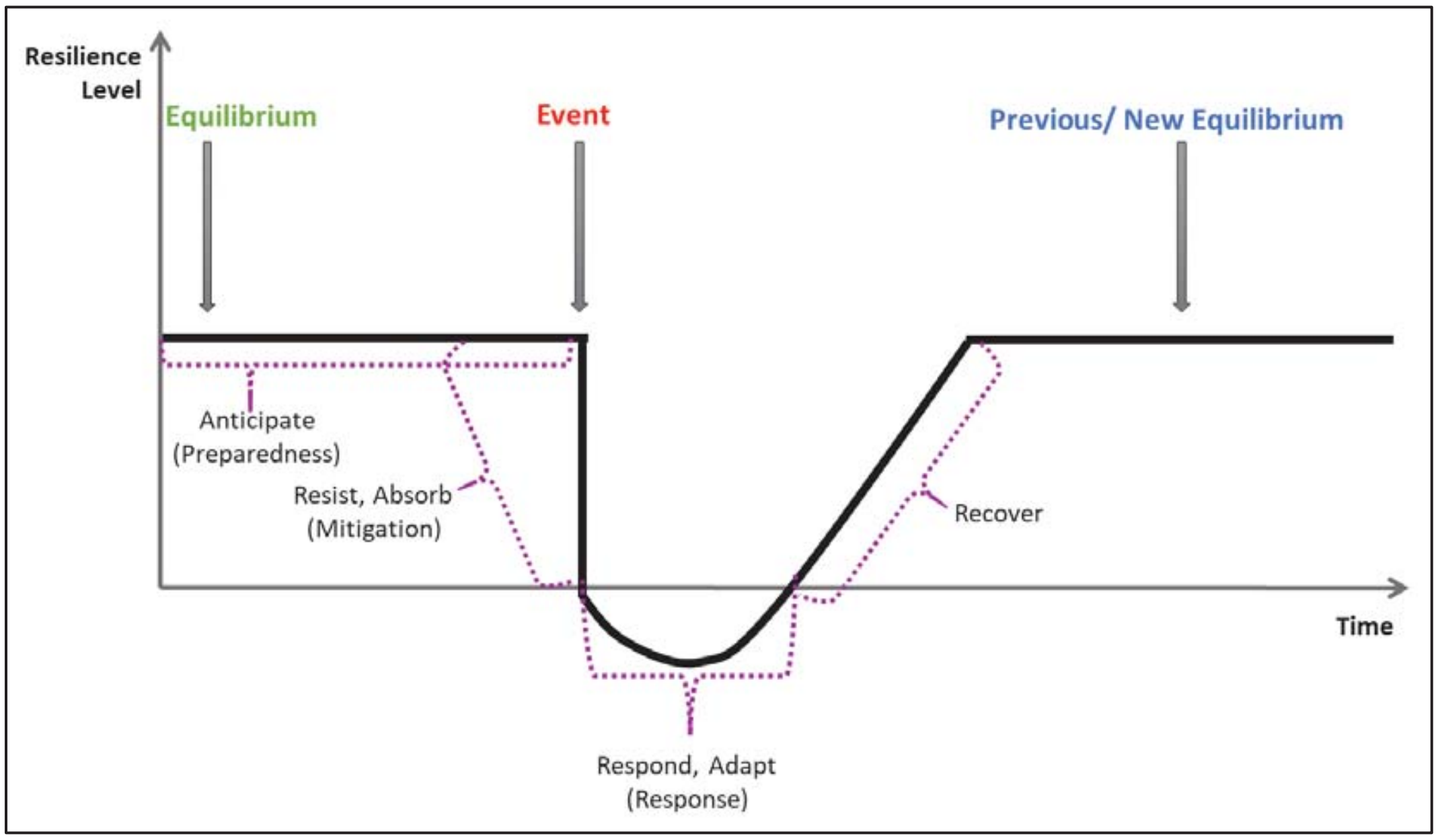

**Figure 2 Phases of resilience [20]**

These four capabilities address the four phases of resilience shown in Figure 2 and form the foundation for the discussion and quantification of microgrid resilience strength. It is also worth noting that *resilience* is associated with the ability to sustain and recover from low-probability high-impact events. In this study, the events (or threats) against which the system's resilience is quantified are external threats, i.e. external perturbances introduced in the energy system as a result of natural disasters, unusual weather conditions, and physical or cyber-attacks. For example, a heatwave causing cooling loads to overload the microgrid beyond expected demand is an example of a low-probability high-impact external threat. A detailed description of the threats considered is provided in [4]. This is differentiated from *reliability,* which focuses on strengthening the system against high-probability low-impact events occurring during normal operations, such as internal component malfunctioning. For example, a non-malicious chance failure of the microgrid falls under the reliability domain. The measures to prevent such events should be addressed during the planning/design process with the goal of maintaining desired reliability levels. Reliability, however, does form one component of the larger resilience equation. For a system to be resilient, having reliable performance is a *necessary* condition but not a *sufficient* one, i.e. being reliable doesn't guarantee that the system will be able to sustain critical load in the face of external threats.

[1] Critical loads are those loads to which power supply has to be maintained under any circumstances.

### *A. Infrastructural and Operational Resilience*

Having established a clear differentiation between the threats pertaining to *resilience* and the disturbances related to *reliability*, we further distinguish between infrastructural and operational resilience. "Infrastructural resilience" encompasses the robustness and redundancy of the hardware components of the microgrid and the resourcefulness of the system to resolve issues with sufficient rapidity to continue operating at normal or near-normal performance levels. "Operational resilience" describes the methods used to operate the microgrid assets in a resilient way to "make the system better able to absorb the impact of an event without losing the capacity of function" [21] These two aspects are highly interdependent, and to keep the critical load 'on', the system must be resilient in both dimensions.

The goal of operational resilience is to ensure the system absorbs the impact of an event through operational strategies like load shedding, and continues to supply the critical load. From a physical perspective, operational resilience requires plans, procedures, and dispatch strategies that allow potentially degraded equipment to keep functioning to serve the critical load. From a cyber perspective, operational resilience requires the analytics and controls software of the microgrid to have the ability to detect a cyber-attack, defend it, block it, and continue supplying the energy as needed. The goal of infrastructural resilience, on the other hand, is to minimize the time needed for the system's hardware and software components to get back to full functional capacity. For the physical layer, this means quickly repairing or replacing damaged components. For the cyber layer, this means securing the communications infrastructure like wired and wireless connections between nodes.

### *B. Resilience Hierarchy*

Though infrastructural and operational resilience address two different aspects of resilience quantification, they are neither *mutually exclusive* nor *independent* dimensions of microgrid resilience. Making a microgrid resilient enough to serve the critical loads during an extreme event requires coordination of both infrastructural and operational aspects. Figure 3 depicts the resilience hierarchy that illustrates the overarching concept of holistic resilience quantification.

When microgrid infrastructure (generation assets, inverters, connecting cables for local distribution, battery storage, controllers) is rendered non-functional by an extreme event, then operational resilience becomes irrelevant because the critical load cannot be served and restoration of the hardware components (in case of physical damage) or the control systems (in case of a cyber-attack) is the primary concern in this scenario. Therefore, prioritizing the hardening of infrastructure (in both physical and cyber dimensions) provides the first line of defense against extreme events. As long as the infrastructure is partially or fully functional however, *operational resilience* becomes of prime importance.

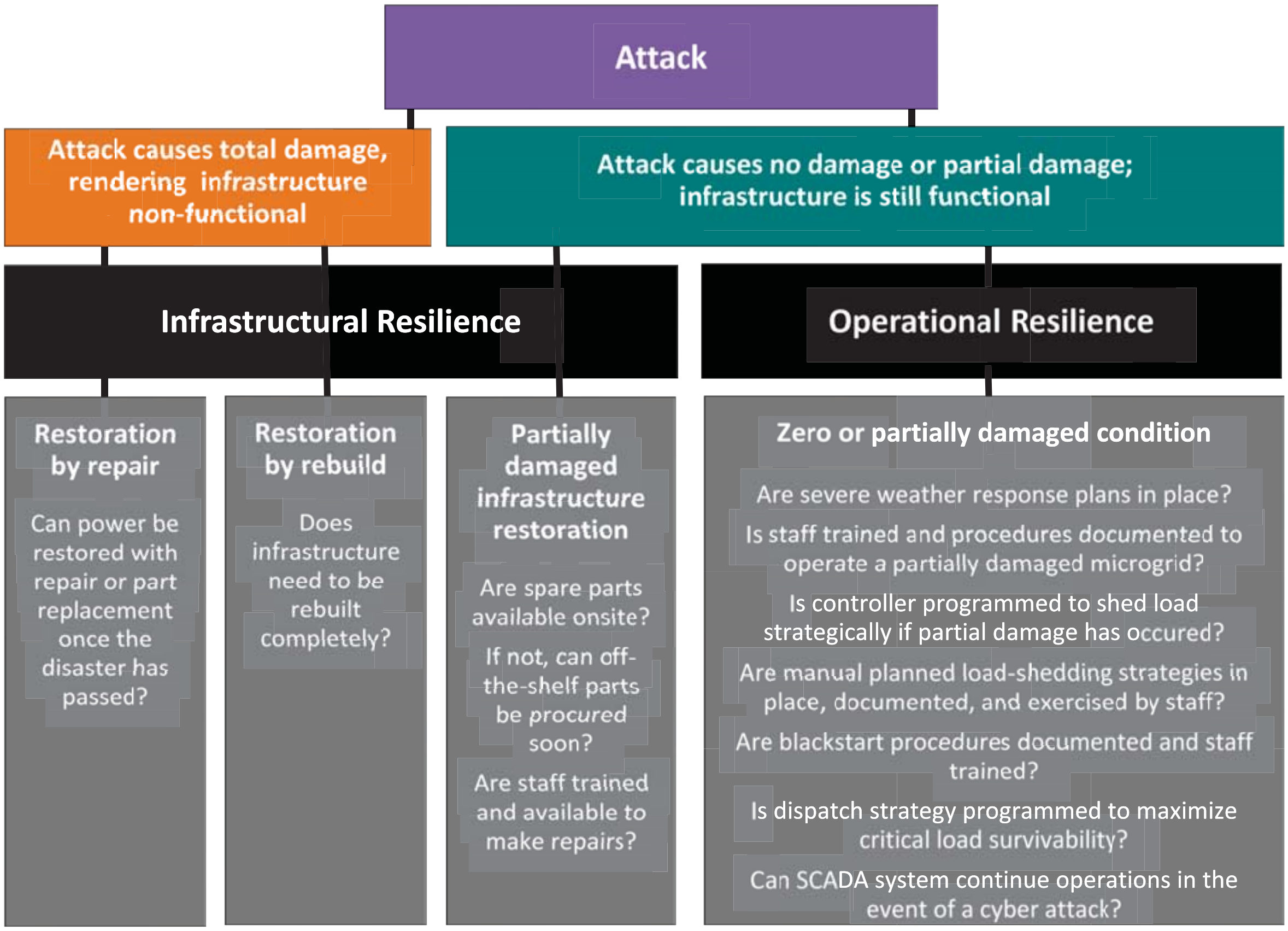

**Figure 3 Proposed Resilience Hierarchy**

## III. Microgrid Resilience Metric – A holistic Assessment Framework

In this section, we propose a metric to to quantify the operational and infrastructural resilience of microgrids considering the *threats to which it is exposed* and the *performance it is expected to deliver* should such adverse events be realized.

### A. *Threats, vulnerabilities, and vulnerability impacts*

We propose a resilience metric that captures the cumulative and weighted effect of *threats*, *vulnerabilities*, and *vulnerability impacts* on both *operational* and *infrastructural* resilience of a microgrid. In the context of this assessment, *threats* are defined as external low-probability high-impact events, either natural or human-caused, with the potential to adversely impact a microgrid's ability to meet the critical load (i.e. hurricanes, terrorist attack). *Threats* are parameterized in terms of their probability of occurrence over a one-year duration at any level of severity, such that 100% probability *threats* are certain to occur every year, and 20% probability *threats* would be expected to happen once every five years. The *threat* probability parameter describes the likelihood of a type of event happening, but it does not describe the impacts expected to occur if the threat is realized. Such impacts are expressed as *vulnerabilities* which are described later in this section. Each *threat* is accompanied by a *level of importance* parameter, which is a scalar between zero and one that is representative of how much the *threat* should contribute to the cumulative metric; zero being not at all and one being of utmost importance. The *level of importance* allows the resilience assessment to be contextualized within the microgrid's constituents' values and objectives. Thus, a modeler may choose to assign a high level of importance to all conceivable *threats*, or, alternatively, amplify, diminish, or nullify the contribution of certain *threats* based on assessment priorities.

Each *threat* will also be associated with one or more *vulnerabilities*, which are direct adverse resilience impacts that occur as *threats* are realized (i.e. high winds take down power lines during a hurricane, saboteur destroys a transformer). *Vulnerabilities* are similarly parameterized in terms of their probability of occurrence should conditions for the *threat* be met. For example, considering the *threat* imposed by a hurricane, the likelihood of a flooding *vulnerability* being exploited may be higher for ground-mounted generation assets than rooftop generators.

Each *vulnerability* in turn has a distinct *vulnerability impact* on both operational resilience and infrastructural resilience. *Vulnerability impacts* on operational resilience are parameterized in terms of the percent of critical load not served during a threatening event. A very high *vulnerability impact* on operational resilience would be one in which the supply of power is completely curtailed. Note, isolated damage to a subset of system components may not be sufficient to degrade operational resilience if the critical load is minimal, or if there is sufficient redundant generation. *Vulnerability impacts* on infrastructural resilience are quantified based on the ratio of restoration costs to the aggregate embedded cost of the system (i.e. an impact that destroys a single generator would reflect the proportional cost of restoring that system component to the capital costs of the total system). Note that restoration costs may include the value of lost load if the operator is liable for such losses. Moreover, while restoration costs may conceivably exceed system capital costs, *vulnerability impacts* are bounded between zero and one. The *vulnerability impact* parameter expresses a relative degree of impact; it need not express precise cost ratios.

The definition of system-level scope for *vulnerabilities* and *vulnerability impacts* (i.e. hurricane winds damage solar PV *vs* hurricane winds damage all generation capacity) is at the discretion of the modeler; what is important is that the *vulnerability impact* reflects the *vulnerability* as defined.

### *B. Parameterization*

The framework we put forth for resilience analysis parametrizes *threat* and *vulnerabilities* based on likelihoods of occurrence, and *vulnerability impacts* on the degree of impact quantified as a percentage. In some cases, such precise numeric values can be estimated from historic trends or advanced site-specific modeling capabilities. We expect though in most cases such data does not exist, is overly burdensome to obtain, or exists within a high degree of uncertainty, especially when considering projections out into the future. A solution widely used in studies of risk analysis is to employ qualitative descriptions in place of quantitative data, often gathered from a group of experts. We begin with a qualitative assessment of parameters and extend these to quantitative ranges by means of the lookup provided in Table 1. Note that such mappings of severity descriptions to quantitative ranges are intended for providing a starting point for analysis; further differentiation of severity categories will help to refine the resolution of results where appropriate. Also, note that parameters may be expressed as groups of ranges (i.e. *Very Low* to *Very High* may describe the vulnerability impact of a single generation asset).

| Classification | Range |
|---|---|
| *Negligible* | 0 – 0.01 |
| *Very Low* | 0.01 – 0.05 |
| *Low* | 0.05 – 0.2 |
| *Moderate* | 0.2 – 0.5 |
| *Considerable* | 0.5 – 0.7 |
| *High* | 0.7 – 0.9 |
| *Very High* | 0.9 - 1 |

**Table 1 Rating Levels and Corresponding Ranges**

Diving deeper into Table 1, we see that a ‘very low’ *threat* would be one that we expect to have a 1-5% rate of occurrence in a year. In other words, we would expect to see this type of *threat* once every 20 to 100 years. Likewise, a ‘very low’ *vulnerability* is one that would only occur rarely if conditions for the *threat* were met. To understand *vulnerability* ratings, consider a simple microgrid consisting of a solar PV panel in an area proximate to mature trees and prone to high winds. While the high wind *threat* may be ‘very high’ if such events are common, the *vulnerability* to physical damage from a fallen tree during high winds may be ‘very low’ given that in only one in a hundred high wind events would wind-induced tree damage be expected to occur.

Building on this example, and to understand the operational resilience of this system, consider that this simple PV system has a series of solar PV panels connected in series to an inverter. In the event that a tree did fall we might expect a ‘very low’ loss of generation capacity if only one panel at the edge of the series is impacted such that only 1-5% of the critical load is lost. However, a strike farther down the series closer may cause a complete loss of critical load. Accordingly, we classify the potential operational resilience impact as ‘very low’ to ‘very high’. With more information about the likelihood of tree damage outcomes on the system, we would be able to say that the distribution of operational impacts is more likely to occur towards the higher (or lower) end of the impact spectrum. Moreover, an intervention that parallelizes panels (or physically hardens this infrastructure) may reduce the importance of any one component such that we drop the worst-case scenario impact on operational resilience to ‘moderate’ or even ‘low’. Finally, to understand infrastructural resilience consider that the microgrid’s vulnerability to physical damage from trees is only expected to damage a subset of panels. Proportional to the capital costs of the total system, such repair or replacement costs may constitute no more than 50%, and with this presumption, we would identify a ‘moderate’ impact on infrastructural resilience.

*C. Monte Carlo Analysis Procedure*

A microgrid's resilience assessment begins with listing all relevant *threats* to a system, inclusive of severe weather events (i.e. thunderstorms), natural disasters (i.e. earthquakes), and human factors (i.e. terrorism). *Threat* likelihoods are parameterized as described above and assigned a *level of importance.* Next, the likelihood that *vulnerabilities* will be realized if a *threat* occurs are contextualized to possible outcomes for system components (i.e. the *threat* of a hurricane may expose the *vulnerability* that PV panels may be damaged, but not pose a risk of damage to a well-protected generator) and parameterized. Likewise, *vulnerability impacts* on operational and infrastructural resilience are parameterized for each *vulnerability.*

Having established quantitative ranges for *threats, vulnerabilities,* and *vulnerability impacts* we employ a Monte Carlo approach to sample an empirical range of potential resilience outcomes for the microgrid. A Monte Carlo approach relying on repeated random sampling to obtain numerical results is used because of the high uncertainty is these values. Randomness is used to solve the problem of obtaining an empirical range of potential resilience outcomes, which are deterministic in principle yet probabilistic in interpretation.

We first assess operational and infrastructural resilience separately. For each identified and parameterized *threat* and *vulnerability* combination in the resilience category (i.e. operational or infrastructural), we compute a residual resilience score as described by the function in Equation 1, shown below, 1 million times. On each iteration, we insert a *threat, vulnerability,* and *vulnerability impact* value at random from within the prescribed bounds of each parameter. The mean of all iterations is then compiled into a metric for the risk.

$$f(l,v,t,i) = \ threat\ level\ of\ importance\ (l)\ x\ probability\ of\ threat(t)\ x$$
$$probability\ vulnerability\ (v)\ x\ impact\ of\ vulnerability(i) \tag{1}$$
$$\forall\ l,v,t,i\ \in (0,1)$$

The resulting mean risk for each *threat* and *vulnerability* combination is then aggregated into an aggregate resilience score as described in Equation 2 below.

$$Resilience\ Score = \in f(l,t,v,i) \tag{2}$$

Finally, after having computed an operational and infrastructural resilience score, we combine the two into a total risk score by means of simple averaging as follows:

$$Risk\ Score\ Total = 1 - \frac{(Risk\ Score\ Operational +\ Risk\ Score\ Operational)}{2}$$

Given the variability in the number and types of potential threats and vulnerabilities to which a microgrid may be exposed, such an empirically derived risk score is not intended to be compared across sites. Instead, this metric is best employed as a baseline for assessing intervention alternatives at the site. Having established a baseline, a modeler employing this methodology can assess the relative effectiveness of mitigations on overall risk by adjusting the initial parameterization then rerunning the analysis.

In the following sections, we provide a case study to illustrate the types of *threats*, *vulnerabilities,* and *vulnerability impacts* one would consider in assessments of microgrid operational and infrastructural resilience. We establish an approximate location and feasible set of generation technologies to make reasonable assumptions about magnitudes and distributions of *threats*, *vulnerabilities,* and *vulnerability impacts.*

## IV. Case Study - Coastal Community in New England

In this case study, we consider a small residential town on the New England coast that has built a grid-tied microgrid to sustain critical services during a larger utility grid outage. In the event of a grid failure, this system will power municipal facilities, medical centers, emergency centers, and food service providers. The microgrid consists of 1 MW rooftop solar PV, a 1 MW wind turbine, 8 MW natural gas generators, 4 MW (8 MWh) battery storage and above-ground distribution lines.

*A. Threat, Vulnerability, and Impact Assessment*

We next describe in detail the *threat*, *vulnerability,* and *vulnerability impact* rankings for this site, which is outlined in Table 2 below and incorporates the qualitative mappings described in Table 1 above. We make reasonable attempts to bound the distribution of each ranking's intensity. However, since this example is intended to demonstrate how to determine a baseline and assess intervention alternatives, rather than determine a site-specific resilience assessment for the specific location, we do not consult expert opinion or attempt to base ranges on published figures. We also assign an importance level of one to all non-negligible threats, meaning we assume that the constituents care equally about all threats that would be expected to occur in the region (i.e. concerned with hurricanes, not tsunamis).

We expect this New England microgrid would be exposed to a moderate to considerable hurricane *threat* (experiencing a hurricane every one to five years). Regarding this *threat*, we identify a set of wind and water-induced *vulnerabilities* which include damage to solar PV, wind turbines, and distribution lines. We also identify a set of *vulnerabilities* resulting from the storage and natural gas generator exposure to coastal flooding. The likelihood of these weaknesses being exploited tends towards a lower degree of severity, given that hurricanes have typically diminished in strength by the time they reach New England. Nevertheless, all such *vulnerabilities* would be expected to induce considerable *vulnerability impacts* on operational and infrastructural resilience in a worst-case scenario where the utility grid has gone down. Distribution line loss resulting from wind and water damages is responsible for the largest individual impacts on resiliency because of the far-reaching impacts of this asset not being able to function at all. During flooding, damage to either storage or generators alone would notably impact resiliency (given the relative capacity of these technologies within the greater system). We retain the possibility that no damage occurs, or that the grid is capable of serving critical loads, by setting the lower bound on resilience impacts to negligible. Moreover, we also recognize that reduced visibility and strong winds resulting from a hurricane are likely to lead to *vulnerability impacts* in the form of PV generation losses and, to a lesser extent, wind generation curtailment. In these cases, operational resilience would experience very low to low degradation, while infrastructural resilience impacts would be negligible.

The concerns posed by hurricanes largely overlap with those of severe winter storms (i.e. blizzards, nor'easters, etc.), which commonly occur in the North East. As such *threats* play out, we estimate conditions are commonly reached that cause solar PV to experience power losses, but that it would be rarer that wind would need to be curtailed on account of high winds. Each of these *vulnerabilities* would in isolation have very low to low *vulnerability impacts* on operational resilience (because of ample generator and storage capacity) and result in no significant *vulnerability impact* on infrastructural resilience. Moreover, both exposed solar PV panels and wind turbines have a low *vulnerability* to damage from a winter storm's snow, ice, and wind. In these cases, both critical load losses and infrastructure damage would be negligible to moderate. In contrast to hurricanes, we assume no operational or infrastructural *vulnerability impacts* resulting from damage to storage or natural gas generators during winter storms.

Our assessment of high winds and severe thunderstorm *threats* at this location also follows the reasoning described for hurricanes and severe winter storms. However, we generally expect thunderstorms to have more mild *vulnerability impacts* on operational and infrastructural resilience. Also, rare damage from lighting events does pose potentially high and moderate *vulnerability impacts* on operational resilience and infrastructural resilience respectively. Damage from lightning may cause widespread outages but is expected to impact isolated components that are readily repaired or replaced.

Hail, too, is a high *threat* in New England. We expect hail in the region to rarely damage solar PV panels, with very low implications for operational and infrastructural resilience.

Situated coastally, the microgrid under study is also conceived to be exposed to low to moderate risk from flooding. During such events, solar PV panels, wind turbines, and distribution infrastructure would be expected to retain high operational and infrastructural resilience, while the natural gas generation and storage assets would experience a potentially moderate vulnerability to flooding. Thus, we would expect partial resilience losses even in worst-case scenarios.

A microgrid in New England would be at a high risk of experiencing an earthquake, though very rarely do earthquakes in this region result in physical damage. Accordingly, the *vulnerability* to physical damage from such *threats* is very low. Still, given the historic magnitudes of earthquakes in this region, considerable *vulnerability impacts* on operational and infrastructural resilience are possible. Similarly, tornados do occur in New England annually, such that the *threat* is high, but the *vulnerability* to any one event is very low. New England tornados can cause severe property damage, and so we recognize potentially considerable *vulnerability impacts* on operational and infrastructural resilience.

This microgrid's electronic components (i.e. control systems, inverters) would furthermore be vulnerable to damage from rare electromagnetic pulses (EMP). For the purposes of this classification and to avoid duplication of vulnerabilities, we consider just those EMP's originating from solar weather, and exclude those originating from lightning or weapons in this grouping in Table 2. EMP's, while rare, have the potential to cause complete outages by damaging electronic equipment which facilitates microgrid operations. Generation assets, however, are expected to stay largely undamaged such that they can come online soon after control systems are replaced. Accordingly, the infrastructural *vulnerability impact* is less serious than the impact on operational resiliency.

Furthermore, it is conceivable (though not necessarily likely) that this microgrid's fossil-fuel-based generation assets would need to be curtailed due to economics in the event natural gas prices spike. While uncertainty exists, current prevailing trends suggest price spike *threats* are low and that few price spike events would make the microgrid vulnerable to generation curtailment. *Vulnerability impacts* could considerably reduce operational resilience in a worst-case scenario but have no impact on infrastructural resilience.

We do not consider wildfires, droughts, and tsunamis to be capable of having more than negligible impacts on the operational and infrastructural resilience of a microgrid in this location. Moreover, due to its small size and a presumed lack of critical government infrastructure in this example community, the *threat* of cyberattack and terrorism is assumed to be very low for a microgrid. All the same, we recognize two levels of cyberattack – a low probability controls override capability that does not damage the system but causes partial to complete outages, and the more rare and sophisticated attack that causes the microgrid to physically damage itself (i.e. overloading lines, overworking batteries). We also recognize that terrorism and sabotage (direct physical damage inflicted on system components) require less sophistication but are potentially highly destructive to the microgrid. Thus, within this classification the range of possible *vulnerability* likelihoods and *vulnerability impacts* is expansive.

| THREAT | THREAT PROBABILITY & IMPORTANCE | VULNERABILITY | VULNERABILITY PROBABILITY | IMPACT ON OPERATIONAL (% CRITICAL LOAD LOST) | IMPACT ON INFRASTRUCTURE (EFFORT OF REPAIRING SYSTEM) |
|---|---|---|---|---|---|
| HURRICANE | **PROBABILITY** *Moderate to Considerable* 0.2 – 0.7 **IMPORTANCE** 1 | Clouds and Rain Lead to PV Generation Losses | *Considerable* *0.5 – 0.7* | *Negligible to Low* 0 – 0.05 | *Negligible* 0 – 0.01 |
| | | High Winds Leads to Turbine Generation Losses | *Low to Moderate* *0.2 – 0.5* | *Negligible to Low* 0 – 0.2 | *Negligible* 0 – 0.01 |
| | | High Winds Damages PV | *Low* 0.05 – 0.2 | *Negligible to Moderate* 0 – 0.5 | *Negligible to Moderate* 0 – 0.5 |
| | | High Winds Damage Turbine | *Low* 0.05 – 0.2 | *Negligible to Moderate* 0 – 0.5 | *Negligible to Moderate* 0 – 0.5 |
| | | High Winds Damage Distribution | *Low* 0.05 – 0.5 | *Negligible to Moderate* 0 – 0.5 | *Negligible to Moderate* 0 – 0.5 |
| | | Heavy Rains/Storm Surge Damages Generator | *Low* 0.05 – 0.2 | *Negligible to Moderate* 0 – 0.5 | *Negligible to Moderate* 0 – 0.5 |
| | | Heavy Rains/ Storm Surge Damages Storage | *Low* 0.05 – 0.2 | *Negligible to Moderate* 0 – 0.5 | *Negligible to Moderate* 0 – 0.5 |
| SEVERE WINTER STORM | **PROBABILITY** *High* 0.7 – 0.9 **IMPORTANCE** 1 | Snow and Ice Lead to PV Generation Losses | *Considerable* *0.5 – 0.7* | *Negligible to Very Low* 0 – 0.05 | *Negligible* 0 – 0.01 |
| | | Snow and Ice Lead to Turbine Generation Losses | *Very Low to Low* 0.01 – 0.2 | *Negligible to Low* 0 – 0.2 | *Negligible* 0 – 0.01 |
| | | Snow, Ice, and Wind Damages PV | *Very Low to Low* 0.01 – 0.2 | *Negligible to Moderate* 0 – 0.5 | *Negligible to Moderate* 0 – 0.5 |
| | | Snow, Ice, and Wind Damages Turbine | *Very Low to Low* 0.01 – 0.2 | *Negligible to Moderate* 0 – 0.5 | *Negligible to Moderate* 0 – 0.5 |
| | | Snow, Ice, and Wind Damages Distribution | *Very Low to Low* 0.01 – 0.5 | *Negligible to Moderate* 0 – 0.5 | *Negligible to Moderate* 0 – 0.5 |
| SEVERE THUNDERSTORM | **PROBABILITY** *High* 0.7 – 0.9 **IMPORTANCE** 1 | Clouds and Rain Lead to PV Generation Losses | *Considerable* *0.5 – 0.7* | *Negligible to Very Low* 0 – 0.5 | *Negligible* 0 – 0.01 |
| | | High Winds Leads to Turbine Generation Losses | *Very Low to Low* *0.01 – 0.2* | *Negligible to Low* 0 – 0.2 | *Negligible* 0 – 0.01 |
| | | High Winds and Rain Damage PV | *Very Low to Low*<br>0.01 – 0.2 | *Negligible to Low*<br>0 – 0.2 | *Negligible to Low*<br>0 – 0.2 |
| | | High Winds and Rain Damage Turbine | *Very Low to Low*<br>0.01 – 0.2 | *Negligible to Low*<br>0 – 0.2 | *Negligible to Low*<br>0 – 0.2 |
| | | High Winds and Rain Damage Distribution | *Very Low to Low*<br>0.01 – 0.5s | *Negligible to Low*<br>0 – 0.2 | *Negligible to Low*<br>0 – 0.2 |
| | | Lightning Causes Electrical System Damage | *Very Low*<br>0.01 – 0.05 | *Negligible to High*<br>0 – 0.9 | *Negligible to Moderate*<br>0 – 0.5 |
| HAIL | **PROBABILITY**<br>*High*<br>0.7 – 0.9<br>**IMPORTANCE**<br>1 | Infrastructure Damage to PV | *Very Low to Low*<br>0.01 – 0.2 | *Negligible to Very Low*<br>0 – 0.5 | *Negligible to Very Low*<br>0 – 0.5 |
| HIGH WIND | **PROBABILITY**<br>*Moderate to Considerable*<br>0.2 – 0.7<br>**IMPORTANCE**<br>1 | High Winds Leads to Wind Generation Losses | *Considerable*<br>*0.5 – 0.7* | *Negligible to Low*<br>0 – 0.2 | *Negligible*<br>0 – 0.01 |
| | | Infrastructure Damage to PV | *Very Low to Low*<br>0.01 – 0.2 | *Negligible to Low*<br>0 – 0.2 | *Negligible to Low*<br>0 – 0.2 |
| | | Infrastructure Damage to Turbine | *Very Low to Low*<br>0.01 – 0.2 | *Negligible to Low*<br>0 – 0.2 | *Negligible to Low*<br>0 – 0.2 |
| | | Infrastructure Damage to Distribution | *Very Low to Low*<br>0.01 – 0.2 | *Negligible to Low*<br>0 – 0.2 | *Negligible to Low*<br>0 – 0.2 |
| FLOODING | **PROBABILITY**<br>*Low to Moderate*<br>0.05 – 0.5<br>**IMPORTANCE**<br>1 | Infrastructure Damage to Generator | *Very Low to Moderate*<br>0.01 – 0.5 | *Negligible to Considerable*<br>0 – 0.7 | *Negligible to Considerable*<br>0 – 0.7 |
| | | Infrastructure Damage to Storage | *Very Low to Moderate*<br>0.01 – 0.5 | *Negligible to Considerable*<br>0 – 0.7 | *Negligible to Considerable*<br>0 – 0.7 |
| EARTHQUAKE | **PROBABILITY**<br>*High*<br>*0.7 – 0.9*<br>**IMPORTANCE**<br>1 | PV Damage | *Very Low*<br>0.01 – 0.05 | *Negligible to Low*<br>0 – 0.2 | *Negligible to Low*<br>0 – 0.2 |
| | | Turbine Damage | *Very Low*<br>0.01 – 0.05 | *Negligible to Low*<br>0 – 0.2 | *Negligible to Low*<br>0 – 0.2 |
| | | Generator Damage | *Very Low*<br>0.01 – 0.05 | *Negligible to Considerable*<br>0 – 0.7 | *Negligible to Considerable*<br>0 – 0.7 |
| | | Storage Damage | *Very Low*<br>0.01 – 0.05 | *Negligible to Considerable*<br>0 – 0.7 | *Negligible to Considerable*<br>0 – 0.7 |
| | | Distribution Damage | *Very Low*<br>0.01 – 0.05 | *Negligible to Considerable*<br>0 – 0.7 | *Negligible to Considerable*<br>0 – 0.7 |
| TORNADO | **PROBABILITY**<br>*High*<br>*0.7 – 0.9* | PV Damage | *Very Low*<br>0.01 – 0.05 | *Negligible to Low*<br>0 – 0.2 | *Negligible to Considerable*<br>0 – 0.7 |
| | **IMPORTANCE**<br>1 | Turbine Damage | *Very Low*<br>0.01 – 0.05 | *Negligible to Low*<br>0 – 0.2 | *Negligible to Considerable*<br>0 – 0.7 |
| | | Generator Damage | *Very Low*<br>0.01 – 0.05 | *Negligible to Considerable*<br>0 – 0.7 | *Negligible to Considerable*<br>0 – 0.7 |
| | | Storage Damage | *Very Low*<br>0.01 – 0.05 | *Negligible to Considerable*<br>0 – 0.7 | *Negligible to Considerable*<br>0 – 0.7 |
| | | Distribution Damage | *Very Low*<br>0.01 – 0.05 | *Negligible to Considerable*<br>0 – 0.7 | *Negligible to Considerable*<br>0 – 0.7 |
| ELECTROMAGNETIC PULSE (non-lightning) | **PROBABILITY**<br>*Very Low*<br>0.01 – 0.5<br>**IMPORTANCE**<br>1 | Inverter Damage | *Very Low*<br>0.01 – 0.5 | *Negligible to High*<br>0 – 0.9 | *Negligible to Moderate*<br>0 – 0.5 |
| FUEL PRICE SPIKES | **PROBABILITY**<br>*Low*<br>0.01 – 0.2<br>**IMPORTANCE**<br>1 | Operation Shutdown | *Very Low to Low*<br>0.01 – 0.2 | *Negligible to Considerable*<br>0 – 0.7 | *Negligible*<br>0 – 0.01 |
| DROUGHT | **PROBABILITY**<br>*Moderate*<br>0.3 – 0.5<br>**IMPORTANCE**<br>1 | - | *Negligible*<br>0 – 0.01 | *Negligible*<br>0 – 0.01 | *Negligible*<br>0 – 0.01 |
| TSUNAMI | **PROBABILITY**<br>*Negligible*<br>0 – 0.01<br>**IMPORTANCE**<br>0 | - | *Negligible*<br>0 – 0.01 | *Negligible*<br>0 – 0.01 | *Negligible*<br>0 – 0.01 |
| WILDFIRE | **PROBABILITY**<br>*Negligible*<br>0 – 0.01<br>**IMPORTANCE**<br>0 | - | *Negligible*<br>0 – 0.01 | *Negligible*<br>0 – 0.01 | *Negligible*<br>0 – 0.01 |
| CYBERATTACK / IT FAULT | **PROBABILITY**<br>*Low*<br>0.05 – 0.3<br>**IMPORTANCE**<br>1 | Controls Override | *Very Low to Low*<br>0.01 – 0.2 | *Negligible to Very High*<br>0 – 1 | *Negligible*<br>0 – 0.01 |
| | | PV Damage | *Very Low*<br>0.01 – 0.05 | *Negligible to Low*<br>0 – 0.2 | *Negligible to Low*<br>0 – 0.2 |
| | | Turbine Damage | *Very Low*<br>0.01 – 0.05 | *Negligible to Low*<br>0 – 0.2 | *Negligible to Low*<br>0 – 0.2 |
| | | Generator Damage | *Very Low*<br>0.01 – 0.05 | *Negligible to Considerable*<br>0 – 0.7 | *Negligible to Considerable*<br>0 – 0.7 |
| | | Storage Damage | *Very Low*<br>0.01 – 0.05 | *Negligible to Considerable*<br>0 – 0.7 | *Negligible to Considerable*<br>0 – 0.7 |
| | | Distribution Damage | *Very Low* 0.01 – 0.05 | *Negligible to Very High* 0 – 1 | *Negligible to Very High* 0 – 1 |
| TERRORISM / SABOTAGE / PHYSICAL FAILURE | **PROBABILITY** *Low* 0.05 – 0.3 **IMPORTANCE** 1 | PV Damage | *Very Low to Very High* 0.01 – 1 | *Negligible to Low* 0 – 0.2 | *Negligible to Low* 0 – 0.2 |
| | | Turbine Damage | *Very Low to Very High* 0.01 – 1 | *Negligible to Low* 0 – 0.2 | *Negligible to Low* 0 – 0.2 |
| | | Generator Damage | *Very Low to Very High* 0.01 – 1 | *Negligible to Considerable* 0 – 0.7 | *Negligible to Considerable* 0 – 0.7 |
| | | Storage Damage | *Very Low to Very High* 0.01 – 1 | *Negligible to Considerable* 0 – 0.7 | *Negligible to Considerable* 0 – 0.7 |
| | | Distribution Damage | *Very Low to Very High* 0.01 – 1 | *Negligible to Very High* 0 – 1 | *Negligible to Very High* 0 – 1 |

**Table 2 Threat, Vulnerability, and Impact Rankings at Coastal New England Microgrid Site**

### *B. Baseline Resilience Calculation*

We compute the average operational and infrastructural residual resilience for each threat across 1 million simulations, before averaging these results again into the cumulative residual risks shown in Figure 4 below. These average resilience metrics provide a baseline for understanding the tendencies of risks at this site.. From Figure 3 we see that variability in residual risk approximates a normal distribution, with operational residual risk (mean 0.0066, standard deviation 0.0013) being somewhat higher than infrastructural risk (mean 0.0053, standard deviation 0.0012). The higher operational residual risk suggests that interventions to reduce operational risks should be prioritized for maximum impact on the site's resilience. Furthermore, taking the cubic root of mean operational and infrastructural residual risk (because the score is essentially derived from the product of three parameters given that level of importance is always 1) suggests that threats, vulnerabilities, and impacts tend towards a 'low' classification.

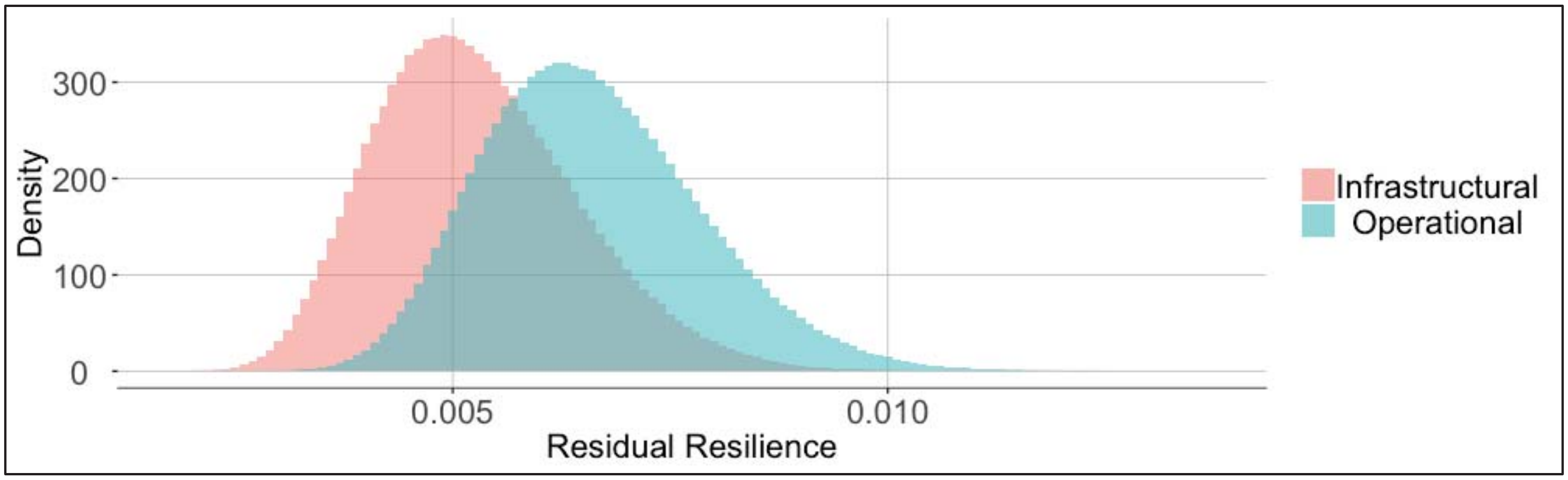

**Figure 4 Operational and Infrastructural Residual Resilience Derived from Monte Carlo Simulation (n=1,000,000)**

When equally weighted and aggregated into a final resilience metric, we yield the distribution shown in Figure 5**.** Resilience values at the New England site range from 0.989 to 0.997 (mean 0.994, standard deviation 0.0008).

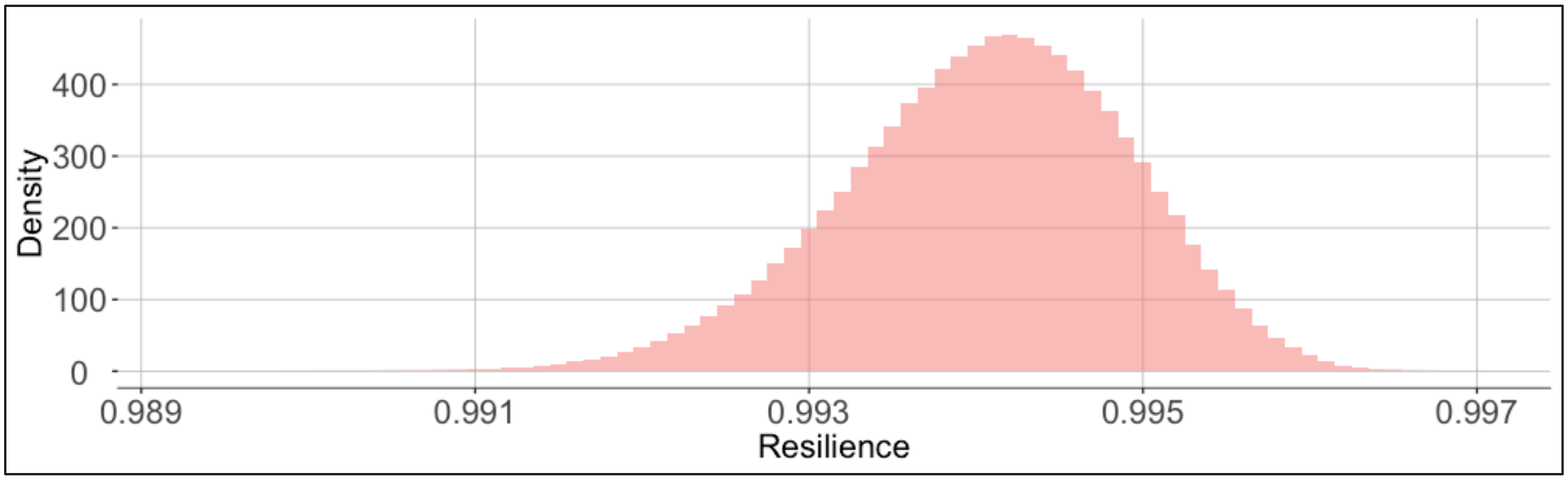

**Figure 5 Resilience Derived from Monte Carlo Simulation (n=1,000,000)**

### *C. Resilience Interventions*

We next explore how the resilience baseline can be used to compare the relative effectiveness of interventions at the site on resiliency. First, we explore the impact of moving all distribution lines underground. This intervention is expected to render obsolete vulnerability of distribution lines to hurricanes, severe winter storms, severe thunderstorms, and high winds. Moreover, this would also cap the vulnerability of terrorism at low, given that the lines would be less readily exposed to attack. However, this intervention would introduce new resilience impacts to flooding and earthquakes.

For comparison, we also examine the physical hardening of generator and storage assets such that they are immune from hurricanes, tornados, earthquakes, and flooding threats (negligible operational and infrastructural impacts). Moreover, these improvements would be expected to reduce vulnerability and resilience impacts resulting from terrorist attacks on storage and generator assets to moderate.

The proposed distribution system intervention reduces the average operational residual risk from 0.0066 to 0.0053 and infrastructural resilience from 0.0053 to 0.0038. Respectively, these reductions represent 20% and 28% reductions in resilience impacts. Overall resilience improves marginally from 0.9938 to 0.9954. Comparatively, the physical hardening of generator and storage assets would be expected to improve operational resilience to 0.0058 (12% reduction), infrastructural resilience to 0.0043 (19% reduction), and overall resilience to 0.9949. Thus, while the interventions are both largely effective relative to baseline conditions, the distribution line enhancements are estimated to more effectively enhance overall resilience. Note this analysis does not answer which intervention is more cost-effective and a more extensive analysis of intervention costs and externalities would be necessary for comprehensive resilience planning.

## V. Summary and Outlook

Microgrids are an important component of power systems resilience. For a microgrid to serve as a resilience resource for the utility grid, the microgrid itself must be resilient enough to absorb, restore, and adapt to the changing circumstances when a low-probability high-impact event occurs. Building resilience within a microgrid requires an assessment of high-risk threats and their potential negative impacts on the microgrid's infrastructure and operations, accounting for the microgrid's vulnerabilities.

The framework laid out in this research provides a methodology for estimating the range and tendency of resilience outcomes. Through Monte Carlo simulation refined to reasonable ranges, we approximate the resilience of a microgrid. As better empirical or modeled data become available, these distributions can be supplemented or replaced to yield more accurate resilience rankings. The framework can also be used to establish a baseline range of resilience outcomes, and quantitatively compare interventions to identify those that have the highest impact on improving the resilience of the microgrid. Therefore, this worklays a solid foundation for conducting resilience analysis during the microgrid planning phase.

The main limitation of this work is that it doesn't consider the economic aspects of resilience intervention deployment. Cost-benefit analysis is standard practice for justifying the business case for building a needed resilience intervention into a microgrid's physical or cyber infrastructure. Future research directions will include quantifying the resilience benefits in monetary terms for inclusion in the cost-benefit analysis.

## VI. Acknowledgment

This work was authored by the National Renewable Energy Laboratory (NREL), operated by Alliance for Sustainable Energy, LLC, for the U.S. Department of Energy (DOE) under Contract No. DE-AC36-08GO28308. This work was supported by the Laboratory

Directed Research and Development (LDRD) Program at NREL. The authors wish to thank Adam Warren (NREL) for providing useful suggestions to refine the manuscript. The views expressed in the article do not necessarily represent the views of the DOE or the U.S. Government. The U.S. Government retains and the publisher, by accepting the article for publication, acknowledges that the U.S. Government retains a nonexclusive, paid-up, irrevocable, worldwide license to publish or reproduce the published form of this work or allow others to do so, for the U.S. Government purposes.